# Folded Periodically-Poled Thin-Film Lithium Niobate for Compact and Efficient Frequency Conversion

**Shengyuan Lu[1], C. J. Xin[1,2], Jiayu Yang[1], Andrea Cordaro[1], Yunxiang Song[1,3], Xinrui Zhu[1], Cornelis A.A. Franken[1,4], Letícia S. Magalhães[1], Matthew Yeh[1,5], Neil Sinclair[1], Amirhassan Shams-Ansari[6], Yaowen Hu[1,7], and Marko Lončar[1]**

[1.] John A. Paulson School of Engineering and Applied Sciences, Harvard University, Cambridge, Massachusetts 02138, USA.

[2.] Present address: HyperLight Corporation, 675 Massachusetts Ave, Cambridge, MA, 02139, USA

[3.] Quantum Science and Engineering, Harvard University, Cambridge, MA 02138, USA.

[4.] Nonlinear Nanophotonics, Department of Science and Technology, MESA+ Institute of Nanotechnology, University of Twente, Enschede, Overijssel, the Netherlands.

[5.] Present address: Center for Quantum Information Physics, Department of Physics, New York University, New York, NY 10003, USA.

[6.] Monarch Quantum, 10150 Meanley Drive, San Diego, CA 92131, USA.

[7.] Present address: State Key Laboratory for Mesoscopic Physics and Frontiers Science Center for Nano-optoelectronics, School of Physics, Peking University, Beijing 100871, China.

Author e-mail address: shengyuan_lu@g.harvard.edu;

[*] Corresponding authors: loncar@g.harvard.edu; shengyuan_lu@g.harvard.edu

## Abstract

Efficient nonlinear frequency conversion is essential for emerging applications ranging from quantum frequency conversion and photon-pair generation to squeezing and integrated optical amplification. Periodically poled thin-film lithium niobate (TF-PPLN) is a leading platform for realizing such processes. However, efficient nonlinear processing typically requires long, straight waveguides that occupy large linear footprints and are susceptible to accumulated phase-matching errors arising from fabrication imperfections. Here, we demonstrate a folded TF-PPLN architecture, consisting of three 1-cm-long straight PPLN sections connected by compact bends with integrated thermo-optic phase shifters. These shifters correct for the phase slips accumulated in bent sections, enabling nonlinear conversion in successive PPLN sections to add constructively. In addition, independent heaters on each PPLN section align their local phase-matching responses to the same target wavelength. Using the combination of wavelength and phase tuning, we achieve a single-pass second-harmonic-generation efficiency of $15777 \pm 1776\ \%/\mathrm{W}$ in a compact 1.35 cm × 0.14 cm area through coherent second-harmonic build-up across multiple physically separated PPLN sections. Furthermore, the optimized nonlinear transfer function of the folded architecture can be thermo-optically tuned while preserving a high conversion efficiency. Using sum-frequency generation to probe the high-conversion regime, we further demonstrate an on-chip photon-conversion efficiency exceeding 60 % at approximately 10 mW on-chip pump power. Such low-power operation could reduce pump-induced noise and relax reliance on wavelength-specific optical amplifiers in quantum-network interfaces, while enabling compact nonlinear circuits for photon-pair generation, upconversion detection, squeezing, and integrated optical parametric processes.

## Introduction

Quantum interconnects are essential for linking heterogeneous quantum nodes over low-loss telecom-fiber networks. Since these nodes generally operate at visible or near-infrared wavelengths, practical quantum networking requires efficient, low-noise frequency conversion between their native optical transitions and the telecom bands[1,2]. Periodically poled thin-film lithium niobate (TF-PPLN) is particularly attractive for this purpose because it combines strong second-order nonlinearity, tight optical confinement, broad optical transparency, and compatibility with integrated photonics. Waveguide-based TF-PPLN in particular allows comparatively broadband and robust operation without requiring simultaneous resonance alignment at multiple wavelengths or carefully engineered coupling rates for efficient photon extraction[3]. In waveguide-based TF-PPLN, increasing the interaction length provides a direct route to reducing the optical power required for high-efficiency conversion, potentially mitigating pump-dependent noise and relaxing the requirements on the pump source. This advantage is especially important when the required pump wavelength lies outside standard optical-amplifier bands, where compact high-power sources may be unavailable.

However, scaling TF-PPLN to longer interaction lengths remains challenging. One limitation is fabrication-related phase-matching inhomogeneity. Variations in film thickness, etch depth, waveguide width, and poling duty cycle can shift the local quasi-phase-matching condition and reduce the coherent interaction length[8]. Adapted poling has been used to tailor the poling period to the measured waveguide geometry[8]. Segmented thermal tuning provides another post-fabrication approach to realign locally detuned phase-matching spectra[9]. These demonstrations motivate extending post-fabrication phase-matching control to substantially longer waveguides. Recent wafer-scale TFLN poling studies show that long, continuously poled regions remain difficult to fabricate[10], as spatial variations make uniform domain inversion increasingly difficult. Although segmented electrodes improve local poling control, they need nearby test devices for pulse optimization, increasing fabrication complexity.

A second scaling constraint is the linear footprint of a single long PPLN waveguide because it requires an uninterrupted region of the wafer and high-quality lithography within that region. This becomes difficult to accommodate in dense photonic circuits, particularly when the chip must also include lasers, modulators, filters or other components. TFLN waveguides support tight optical confinement and compact bends: reported examples include a 200 μm bending radius in TFLN[11] and a 30 mm ring radius in conventional Ti:$LiNbO_3$ waveguides[12]. These compact bends allow long waveguide paths to be routed within a small chip area. While folded paths have enabled nearly one-meter-long electro-optic modulators within millimeter-scale footprints and low drive voltage[13], here we use this layout freedom for coherently cascaded nonlinear interactions.

Here, we demonstrate a folded TF-PPLN device consisting of three 1-cm-long PPLN sections connected by unpoled waveguide bends. Integrated heaters independently tune the quasi-phase-matching wavelength of each PPLN section, while heaters integrated within bends control the relative phase between the fundamental and second-harmonic fields. Through optimization of these degrees of freedom, we show coherent addition of the fields generated in the three sections producing a normalized SHG of $15777 \pm 1776\ \%/\mathrm{W}$. We further demonstrate thermo-optic wavelength trimming of the optimized transfer function and greater than 60 % on-chip photon conversion in sum-frequency generation at approximately 10 mW on the calibrated pump-power axis. This architecture provides a route toward compact and low-input power frequency converters for interfacing heterogeneous quantum nodes with telecom-fibre networks, while improving layout flexibility, integration density, and fabrication yield.

## Device concept and layout

Conventional proton exchanged, Ti-infused PPLN and TF-PPLN devices are typically implemented as straight waveguides, as illustrated in Fig. 1(a)(i). In this geometry, the achievable nonlinear interaction length is directly constrained by the wafer or die size. For example, a straight waveguide can extend to approximately 10 cm across a representative 4-inch wafer, but such a device would sample thickness variations of approximately 30 nm. In practical wafer-scale fabrication, however, devices are more likely to be arranged within individual dies as schematically shown in Fig. 1(a)(ii), thus limiting the linear footprint available for a single TF-PPLN waveguide. The finite exposure fields of lithography tools impose an additional layout constraint; for example, the ASML TWINSCAN NXE:3400B scanner specifies a maximum exposure field of 26 mm × 33 mm[52]. Hence, we propose a folded geometry as shown in Fig. 1(b), and this enables longer effective interaction lengths while maintaining a reduced linear device footprint. In addition, keeping the PPLN sections within the same region of the wafer helps mitigate the range of film thickness and etch depth variations, as well as poling consistency.

Under ideal conditions, each PPLN section has the same normalized conversion efficiency (CE), $\eta_1 = \eta_2 = \eta_3 = \eta$. If N identical sections are phase matched and their generated SH fields add coherently, the SH amplitude scales linearly with N, and N coherently folded PPLN sections provide an ideal efficiency of $N^2\eta$. Another key requirement for coherent cascading is phase control between adjacent PPLN sections. As the fundamental harmonic (FH) and second harmonic (SH) fields propagate through the connecting waveguides and bends, they accumulate a relative phase due to the difference between their effective indices, $n_{eff,FH}$ and $n_{eff,SH}$. If the relative FH–SH phase accumulated, or phase slip, is not compensated, the SH field generated in one PPLN section can interfere destructively with the SH field generated in the next section, reducing the total CE. Therefore, thermo-optic phase shifters are placed between adjacent PPLN sections to tune the relative FH–SH phase. In addition, each PPLN section is equipped with a peak-shifting thermal tuner to align its phase-matching wavelength to the target operating wavelength.

Fig. 1(c) shows an optical microscope image of a fabricated folded TF-PPLN device. The device consists of three 1.0-cm-long PPLN sections connected by compact bends, integrated into a footprint as small as 1.35 cm × 0.14 cm. The separation between adjacent PPLN sections is 440 μm, and integrated heaters are included for both phase-matching alignment and inter-PPLN phase compensation. Detailed heater dimensions and electrical parameters are provided in the Methods section.

## Device principle and simulation

To evaluate the benefit and practical limits of the folded geometry, we first simulated second-harmonic generation in a single TF-PPLN section. As shown in Fig. 2(a) (i), for a normalized CE of 3200 %/W/cm$^2$, a longer PPLN interaction length reduces the pump power required to reach absolute conversion efficiencies above 95 %. However, increasing the PPLN length also narrows the phase-matching bandwidth. Fig. 2(a)(i) shows that the FWHM decreases from 0.266 nm for 1 cm to 0.089 nm for 3 cm and 0.030 nm for a 9-cm continuous waveguide. Even small uncorrected errors from imperfect metrology, fabrication variations, or poling deviations can degrade coherent build-up in an ultra-long continuous PPLN waveguide. To show that maintaining efficient conversion in a long continuous TF-PPLN requires increasingly detailed thickness measurement, we simulate adapted poling using a 16-mm-long TF-PPLN thickness profile measured at 200-μm intervals[8]. As shown in Fig. 2(a)(ii), increasing the sampling interval from 200 to 1000 μm reduces the CE more prominently for longer devices due to accumulated phase error. Dividing the total interaction length into shorter PPLN sections reduces this sensitivity because each section can be individually aligned after fabrication using local thermo-optic tuning. Figure 2(b) illustrates this tuning principle for three PPLN sections. Before tuning, fabrication-induced variations shift the phase-matching responses of the individual sections. TF-PPLN peak-shifting heaters first align the three responses to a common target wavelength, and assuming no phase-mismatch or loss at the bend, produce a coherent sum that ideally reaches nine times the efficiency of a single section.

In a folded device, the connecting bends introduce two effects: phase slip between the FH and SH waves and optical loss. The inter-PPLN phase mismatch between the FH and SH fields can be written as $\Delta\phi = (2\beta_{FH} - \beta_{SH})\, L_{bend} \mod 2\pi$, where $\beta_j = 2\pi\, n_{eff,j}/\lambda_j, j = \mathrm{FH}, \mathrm{SH}$, and $L_{bend}$ is the effective connecting path length between two PPLN sections. For simplicity, we assume that the FH and SH effective indices are unchanged between the connecting waveguides. Although curvature can perturb the two effective indices differently in practice, the resulting additional phase slip can be compensated by the inter-PPLN phase heater. As shown in Fig. 2(c)(i), a small path-length deviation $\Delta L$ can lead to a large change in $\Delta\phi$. This phase sensitivity directly affects the coherent addition of the SH fields generated in different PPLN sections. Fig. 2(c)(ii) and 2(c)(iii) show the calculated overall CE for two- and three-section folded devices as a function of the connecting path-length deviation. When phase slip increases, destructive interference rapidly suppresses the CE, whereas when the inter-PPLN phase is properly matched, the response approaches the lossless coherent limits with $\eta_2/\eta_1 = 4$ and $\eta_3/\eta_1 = 9$.

Bend loss is characterized by fitting the round-trip loss of racetrack resonators with different straight connecting lengths fabricated using the same process as the final folded PPLN device (Device fabrication section in Methods). The bend loss is extracted from the intercept of the fitted loss versus straight-section length at zero straight-waveguide length, with the intercept divided by two to account for the two bends in each device. More details are explained in the Bend loss and Propagation Loss section in Methods. From Extended Fig. 3, this yields $\alpha_{bend,FH} = 0.05 \pm 0.01$ dB/bend at 1600 nm and $\alpha_{bend,SH} = 0.28 \pm 0.02$ dB/bend at 780 nm for the bend geometry. These values indicate that the additional bend loss is modest and does not fundamentally prevent coherent cascading of nonlinear conversion across multiple PPLN sections.

In addition to bend losses, we also evaluate the impact of propagation loss on the folded device performance. From measurements of the aforementioned racetrack resonators, the extracted propagation loss is $\alpha_{prop,FH} = 0.18 \pm 0.07$ dB/cm at 1600 nm and $\alpha_{prop,SH} = 1.30 \pm 0.23$ dB/cm at 780 nm. Suboptimal oxide quality may contribute to the propagation loss, and can be improved as demonstrated in the prior work[15]. Fig. 2(d) shows the simulated CE for a three-section folded device as a function of the FH and SH propagation losses, with $L_{PPLN} = 3$ cm and a fixed on-chip input power of $P_{in,FH} = 15$ mW. CE decreases as either FH or SH loss increases, with SH loss having a particularly strong impact because generated SH power must propagate through the remaining PPLN sections. The star marks the loss values measured in this work, while the diamond indicates a representative low-loss target. Reported telecom losses down to 1.3 dB/m and visible losses down to 6 dB/m suggest that further scaling should be feasible with optimized TFLN fabrication[11,16].

**Optimization and thermal tuning of a three-section folded TF-PPLN device**

We next experimentally characterize a three-section folded TF-PPLN device. First, the overall normalized CE of each PPLN section is measured. As shown in Fig. 3(a)(i)–(iii), the selected section (red trace) remains near its intrinsic phase-matching condition, while the other two sections (dashed gray traces) are thermally red-shifted to spectrally separate their SH contributions. The multi-peak spectra may reflect local phase-matching inhomogeneity and residual generation in the detuned sections; these measurements do not isolate the contribution of each mechanism. The three PPLN sections have different peak wavelengths and conversion efficiencies, despite the use of adaptive poling. From these isolated spectra, we extract the individual section efficiencies and use them to estimate the expected folded response under ideal spectral alignment and coherent phase addition. The detailed computation is explained in the Methods section. As shown in Fig.3(a)(iv), the simulated maximum overall CE of the three-section device is 17689 %/W and unlike a lossless extrapolation, this efficiency includes the propagation and connection losses of the full three-section device.

After characterization of individual PPLN sections, we optimize the full transfer function using iterative thermal tuning. We choose the longest-wavelength isolated phase-matching peak as the target wavelength, and red-shift other sections through thermo-optic tuning. In parallel, the inter-PPLN phase shifters are adjusted to compensate for phase slips between adjacent PPLN sections. This phase optimization is repeated after peak tuning because changing the PPLN heater powers also modifies the optical phase accumulated. The detailed tuning optimization is explained in the Methods section. Fig. 3(b) shows the evolution of the measured overall conversion-efficiency spectrum during the optimization. The tuning trajectory is shown in Fig. 3(b)(ii) and normalized CE at selected steps is shown in Fig. 3(b)(i). Starting from the initially detuned state in the first panel, the spectrum gradually evolves toward a single dominant peak. The iterative tuning was stopped when additional adjustments no longer increased the peak CE within the experimental uncertainty. The peak CE increases from approximately 10696 %/W before optimization to a value of $15777 \pm 1776$ %/W after 22 tuning steps, and the thermal powers applied are in Extended Table 1. The inferred lossless-equivalent section efficiencies average 3863 % $W^{-1}$ $cm^{-2}$. Coherent addition of these three unequal section amplitudes gives an ideal zero-loss model efficiency of approximately 34392 %/W, about 8.9 times the mean 1-cm section efficiency. This is a theoretical comparison, not a loss-corrected measurement of the optimized device. The full derivation of these uncertainties and the exact zero-loss cascade efficiency is given in the Supplementary Information.

The measured peak overall efficiency is 89 % of the simulated maximum overall CE, corresponding to 94 % of the predicted output-field amplitude. Since we observed negligible crosstalk and small peak-shifts for heater powers below 200 mW (Extended Fig.1), the remaining discrepancy between the optimized experiment result and simulated result is likely due to imperfect heater optimization. Current strategy of sequential scanning effectively performs coordinate-descent optimization and may converge to a local rather than global maximum. Future implementations could adopt better tuning strategies for multi-parameter optimization, such as joint phase-shifter scans[17], progressive local-feedback self-configuration[18], and self-calibration under thermal cross-talk[19,20] if the power applied exceeds 200 mW.

After optimizing the folded device for maximum SHG efficiency, we examine wavelength tuning of the optimized transfer function. This is important because practical frequency converters often need to be tuned to compensate for fabrication offsets to access a target operating wavelength. Starting from the optimized SHG condition in Fig. 3 and based on thermal tuning efficiency measured in Extended Fig.2, we apply additional thermal power to the three PPLN peak-shifting heaters to red-shift their phase-matching spectra together, and re-optimize the phase shifters in bent regions to restore coherent add up. Fig. 3(c)(ii) shows the measured peak SH wavelength and peak CE at different thermo-optic tuning settings. The folded phase-matching peak can be continuously red-shifted while maintaining a similar spectral profile and nearly unchanged peak CE. The average power applied to the PPLN peak-shifting heaters increases with the SH wavelength detuning [Fig. 3(c)(i)]. In contrast, the average power applied to the inter-PPLN phase shifters does not necessarily increase monotonically [Fig. 3(c)(i)], since these heaters set the relative FH–SH phase modulo 2π, allowing each heater to settle at the nearest constructive-interference condition.

**Sum-frequency generation (SFG) and potential applications in quantum networks**

Next, we explore the performance of our device with sum-frequency generation in the high pump power regime, since these conversion processes are needed for quantum network applications. Here, we drive the device using two telecom-band input waves at $\lambda_{\text{pump}} = 1591.9$ nm and $\lambda_{\text{idler}} = 1450.5$ nm, allowing the use of accessible telecom-band laser sources. The two wavelengths generate an SFG signal near $\lambda_{\text{vis}} = 758.96$ nm and the experimental setup is shown in Extended Fig.5. Figure 4 shows the measured on-chip SFG photon CE as a function of input pump power, defined as $\frac{P_{\text{vis,on-chip}}}{P_{\text{idler,on-chip}}}\frac{\lambda_{\text{vis}}}{\lambda_{\text{idler}}} \times 100\ \%$. At 9.97 mW on the current calibrated pump-power axis, the measured photon conversion is 61.5 %, whereas a prior PPLN converter reached approximately 65 % internal conversion at approximately 130 mW internal pump power[21]. The fit gives a normalized power-conversion efficiency of 24,078 ± 656 %/W, accounting for propagation and bend losses in the device. The derivation and uncertainties are given in the Methods. Differences between the fitted curve and the measurements may arise from wavelength-dependent losses, calibration errors, unequal section coupling or residual inter-section phase. The losses were characterized near 1600 and 780 nm rather than at all three SFG wavelengths, so the present data do not uniquely identify the source of the residuals. These results demonstrate that the folded architecture provides sufficient nonlinear interaction strength for efficient wavelength conversion at low pump power, a capability particularly valuable for quantum interconnects. As shown in Figure 4(b), interfacing 1350- or 1550-nm telecom photons with the optical transitions of heterogeneous quantum nodes, including solid-state emitters[21–26], rare-earth-ion memories[27–29], trapped ions[30–33], and neutral atoms[34,35], requires pump wavelengths spanning from the visible to beyond 2 μm.

Since high-power optical amplifiers are not widely available across all wavelengths, the long effective interaction length of folded TF-PPLN would enable efficient, low-pump-power conversion using compact lasers without external amplification.

**Summary and Outlook:**

In conclusion, we have demonstrated a folded TF-PPLN architecture that combines a long effective nonlinear interaction length within a compact footprint. Three 1-cm-long PPLN sections are connected by folded waveguides and independently controlled using quasi-phase-matching heaters and inter-PPLN phase shifters, enabling coherent addition of the generated second-harmonic fields. The optimized device achieves an overall SHG efficiency of $15777 \pm 1776$ %/W, corresponding to 89 % of the predicted efficiency under optimal phase alignment, or 94 % of the corresponding output-field amplitude. The measured efficiency includes the propagation losses together with the bend losses accumulated throughout the folded circuit. The SH output can also be thermo-optically shifted while maintaining a similar spectral profile and peak efficiency. In the high-conversion regime, nondegenerate sum-frequency generation yields an on-chip photon CE of 61.5 % at 9.97 mW on-chip pump power. The five heaters consume a total electrical power of 107.9 mW at the optimized operating point. This static tuning power could be reduced through improved thermal isolation, for example by using a thinner buried-oxide layer.

This experiment demonstrates that spatially separated traveling-wave nonlinear interactions can be composed into a single effective interaction when the quasi-phase-matching condition of each section and the relative phases between sections are independently controlled. The compact bends available in thin-film lithium niobate decouple the total nonlinear interaction length from the longest physical dimension of the circuit, while maintaining coherent field accumulation across the separated sections through active spectral and phase control. This folded architecture may also improve functional yield and robustness: in a continuous ultra-long PPLN waveguide, a local waveguide or poling defect can perturb the nonlinear response of the entire device, whereas shorter sections can be characterized and tuned individually, potentially reducing the impact of local nonuniformities.

The folded architecture is applicable beyond the specific implementation in TF-PPLN demonstrated here. The same approach could be transferred to other thin-film nonlinear platforms, such as thin-film lithium tantalate[36,37], where a longer effective interaction length may help compensate for lower second-order nonlinear susceptibilities. The coherent-cascading principle is also potentially applicable to SPDC[38,39], OPA[40], on-chip frequency-comb self-referencing[42], single-photon upconversion detection[43], squeezed-light generation[41,44], supercontinuum generation[51], and integrated visible or ultraviolet light sources[45,46], all of which could potentially benefit from long, low-loss nonlinear interactions. By independently tuning different sections to distinct quasi-phase-matched processes, folded devices could ultimately support reconfigurable nonlinear photonic circuits[47] that integrate multiple wavelength-conversion and quantum-optical functions within one device.

Furthermore, the folded PPLN architecture provides a nonlinear photonic building block for wavelength-flexible, low-power, and scalable frequency conversion. This capability is particularly relevant for quantum frequency conversion between atomic or solid-state quantum emitters and telecom wavelengths, where pump wavelengths may lie outside standard C/L-band amplification windows and where broadband high-power amplification can be costly or unavailable. With improved fiber-to-chip coupling[48], the required pump power could be supplied directly by compact distributed-feedback (DFB) lasers[4] or integrated semiconductor laser sources[5,6,49] without off-chip optical amplifiers, with the latter approach already demonstrated through the integration of III–V lasers with TF-PPLN for telecom-to-visible light generation[6]. Together with wavelength-accurate PPLN fabrication, folded TF-PPLN devices could support multiplexed interfaces across disparate quantum systems, paving the way for low-loss fiber interconnects for heterogeneous quantum memories.

## Figures

**Figure 1: Folded TF-PPLN waveguide architecture for overcoming wafer-and-die-size limits on interaction length.**

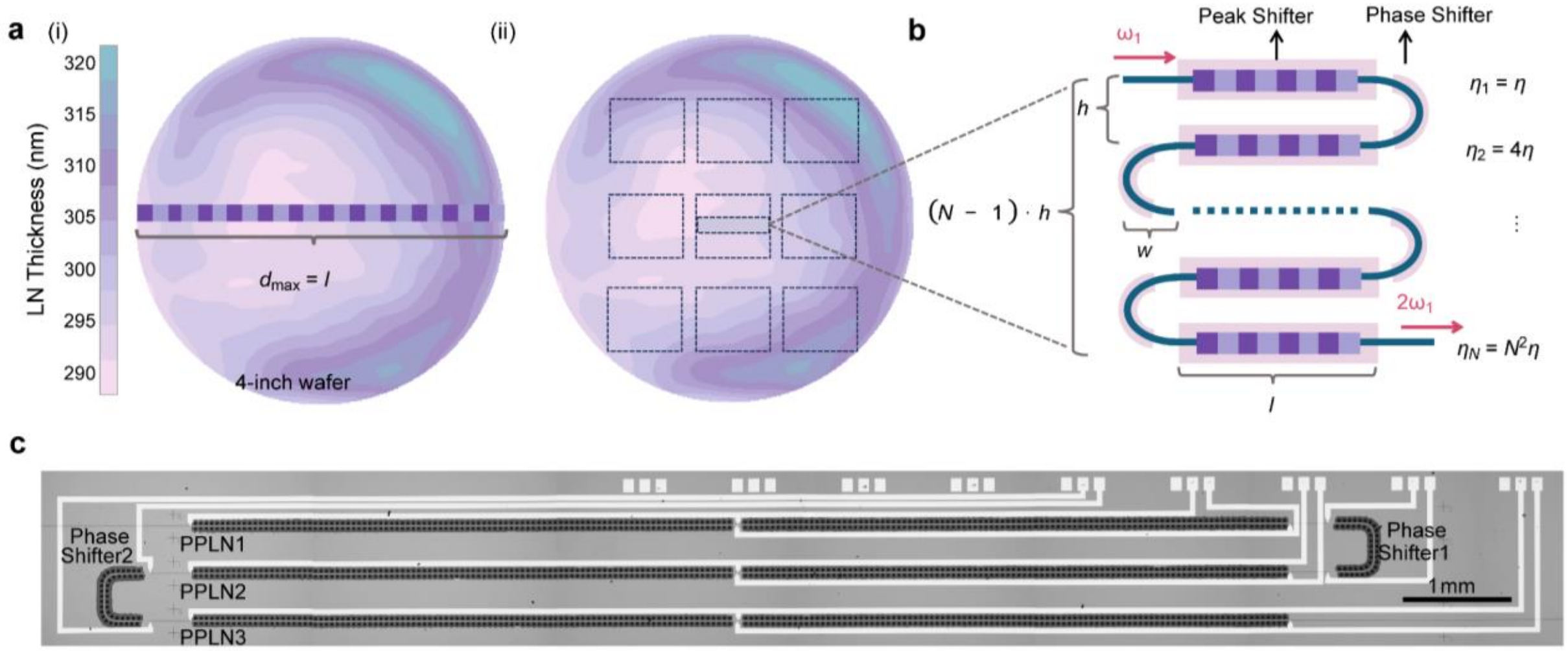

**(a),** Illustration of TF-PPLN layouts overlaid on a representative wafer-scale thin-film thickness map. **(i)** A conventional straight TF-PPLN waveguide, whose maximum periodically poled length $d_{max} = l$ spans a relatively large thickness variation across the wafer. **(ii)** An example layout illustrating how multiple compact folded PPLN devices can be arranged across nine dies on a 4-inch wafer; the reduced device footprint limits the thickness variation sampled by each device. **(b)**, Proposed design of the folded TF-PPLN waveguide for second harmonic generation with input $\omega_{FH}$ and output $\omega_{SH} = 2\omega_{FH}$. The device consists of N TF-PPLN sections, each of length l and the total length is $N \cdot l$. Each PPLN section includes a peak-shifting heater, with a phase shifter placed between adjacent PPLN sections, and the lateral spacing between adjacent PPLN is w. In the ideal limit of equal amplitudes, zero loss, and perfect relative phase alignment at bends, the efficiencies of 1,2, … N coherently folded sections are $\eta, 4\eta \ldots N^2 \cdot \eta$. **(c)**, Optical microscope image of a folded PPLN device, consisting of three PPLN segments each 1.0-cm-long. The vertical distance between PPLN sections is 440 μm. Labels identify the three PPLN heaters and two inter-PPLN phase heaters.

**Figure 2. Design considerations for coherent folded PPLN frequency conversion.**

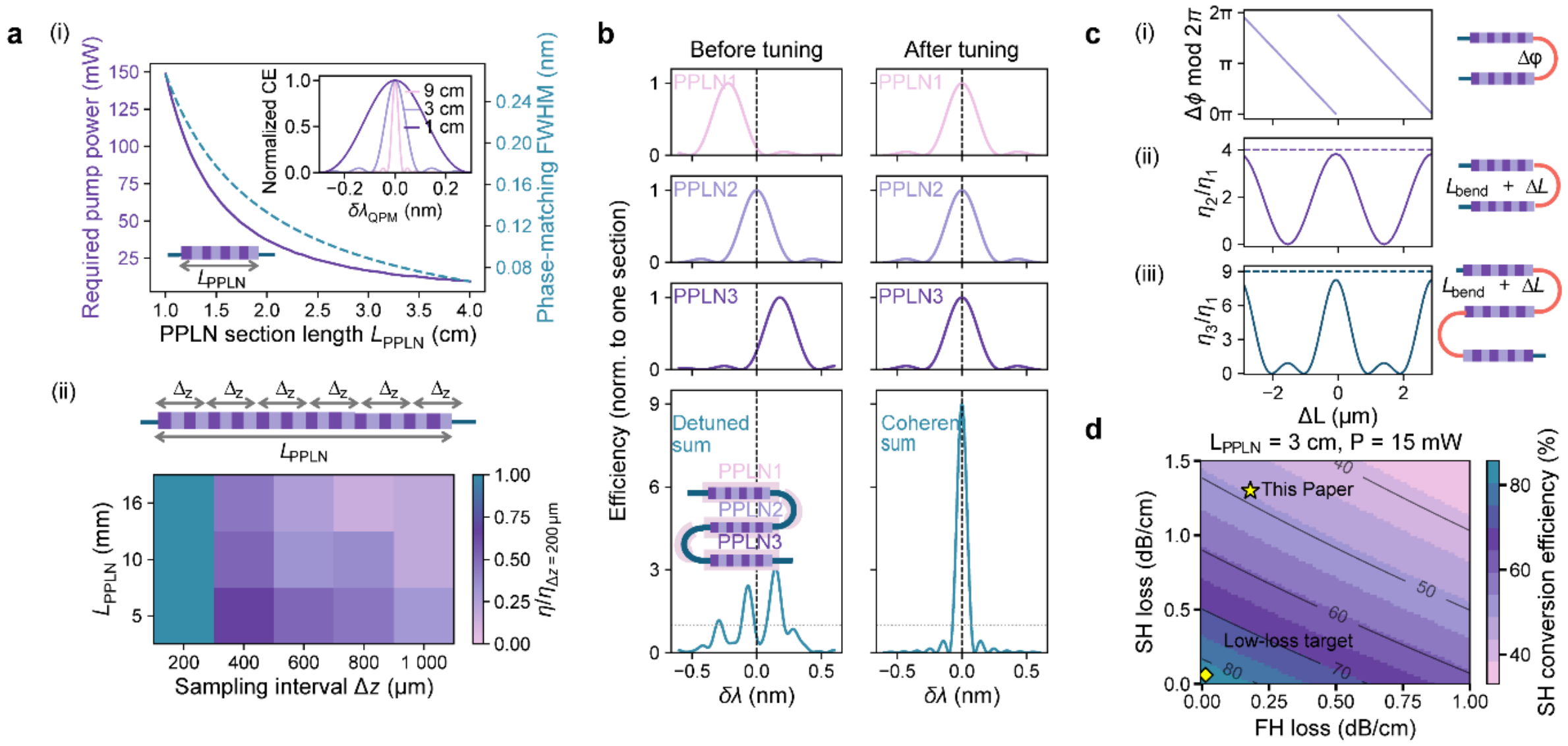

**a**, **(i)** Simulated pump-power required for absolute conversion above 95 % and phase-matching full width at half maximum (FWHM) versus individual PPLN section length, $L_{PPLN}$. The inset shows normalized SHG spectra for 1-, 3-, and 9-cm sections, illustrating the bandwidth narrowing for longer interaction lengths. **(ii)** A 16-mm-long TF-PPLN thickness profile, measured at 200-μm intervals, is used to simulate metrology sampling intervals Δz ranging from 200 to 1000 μm, and used to design adapted poling for different TF-PPLN lengths. The color represents the optimized peak SHG efficiency normalized to that of an ideal phase-matched TF-PPLN. **b**, Schematic of spectral alignment in a three-section folded PPLN. Before tuning, the individual PPLN phase-matching peaks are misaligned, leading to a nonconstructive output. After independent peak tuning and inter-PPLN phase compensation, the SH fields add coherently. **c**, Simulated inter-PPLN phase sensitivity and coherent-efficiency scaling. **(i)** Phase accumulation between FH and SH, Δϕ mod 2π, versus path-length deviation ΔL. **(ii, iii)** Overall SHG efficiency for two- and three-section folded devices versus ΔL, using measured bend losses of 0.05 dB (FH) and 0.28 dB (SH), zero propagation loss, and dashed lines showing the lossless coherent limits $\eta_2/\eta_1 = 4$ and $\eta_3/\eta_1 = 9$. d, Calculated SH CE for a three-section folded device as a function of FH and SH propagation losses, with $L_{PPLN} = 3$ cm and $P_{in,FH} = 15$ mW. The star marks the loss values measured in this work, while the diamond indicates a representative low-loss target.

**Figure 3: Spectral alignment and optimization of a three-section folded TF-PPLN device.**

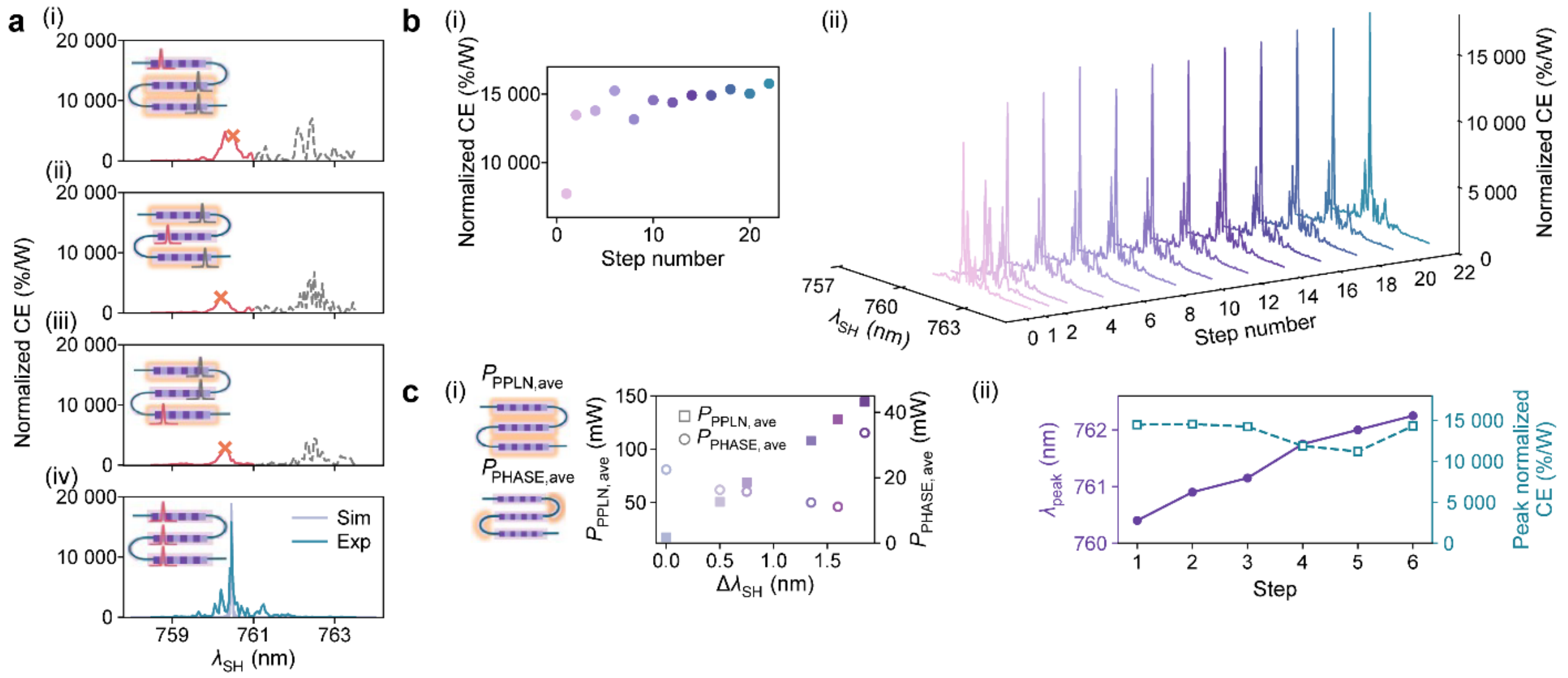

**a**, Measured normalized second-harmonic CE of the individual PPLN sections in the folded device at 1520 nm. **(i)** Response of PPLN1 (red trace), measured by detuning PPLN2 and PPLN3 (dashed traces) with heater powers $P_{PPLN2,PPLN3} = 160$ mW, while keeping $P_{PPLN1} = 0$ mW, **(ii)** Response of PPLN2, measured by detuning PPLN1 and PPLN3 with heater powers $P_{PPLN1,PPLN3} = 160$ mW, while keeping $P_{PPLN2} = 0$ mW, **(iii)** Response of PPLN3, measured by detuning PPLN1 and PPLN2 with heater powers $P_{PPLN1,PPLN2} = 160$ mW, while keeping $P_{PPLN3} = 0$ mW. **(iv)** Simulated overall CE of the three-section folded device assuming perfect spectral alignment and coherent phase matching between all PPLN sections while accounting for the propagation and bend losses in the small-signal limit. **b**, Iterative tuning of the folded TF-PPLN transfer function. **(i)** Peak overall CE at selected steps. **(ii)** Evolution of the measured transfer function at selected tuning steps, showing an increased peak CE as the phase-matching spectra of the individual PPLN sections are aligned and the inter-PPLN phases are optimized. **c**, Thermo-optic red-tuning of the optimized folded SHG spectrum. **(i)** Average heater powers as functions of SH-wavelength detuning, $\Delta\lambda_{SH}$. The left axis shows the average power applied to the three peak-shifting PPLN heaters, while the right axis shows the average power applied to the two inter-PPLN phase-shifting heaters. **(ii)** Measured peak SH wavelength, $\lambda_{peak}$, and peak CE under different heater settings, showing tunable phase matching with preserved peak efficiency.

**Figure 4: Low-input-power sum-frequency generation conversion and pump-wavelength requirements for heterogeneous quantum interconnects.**

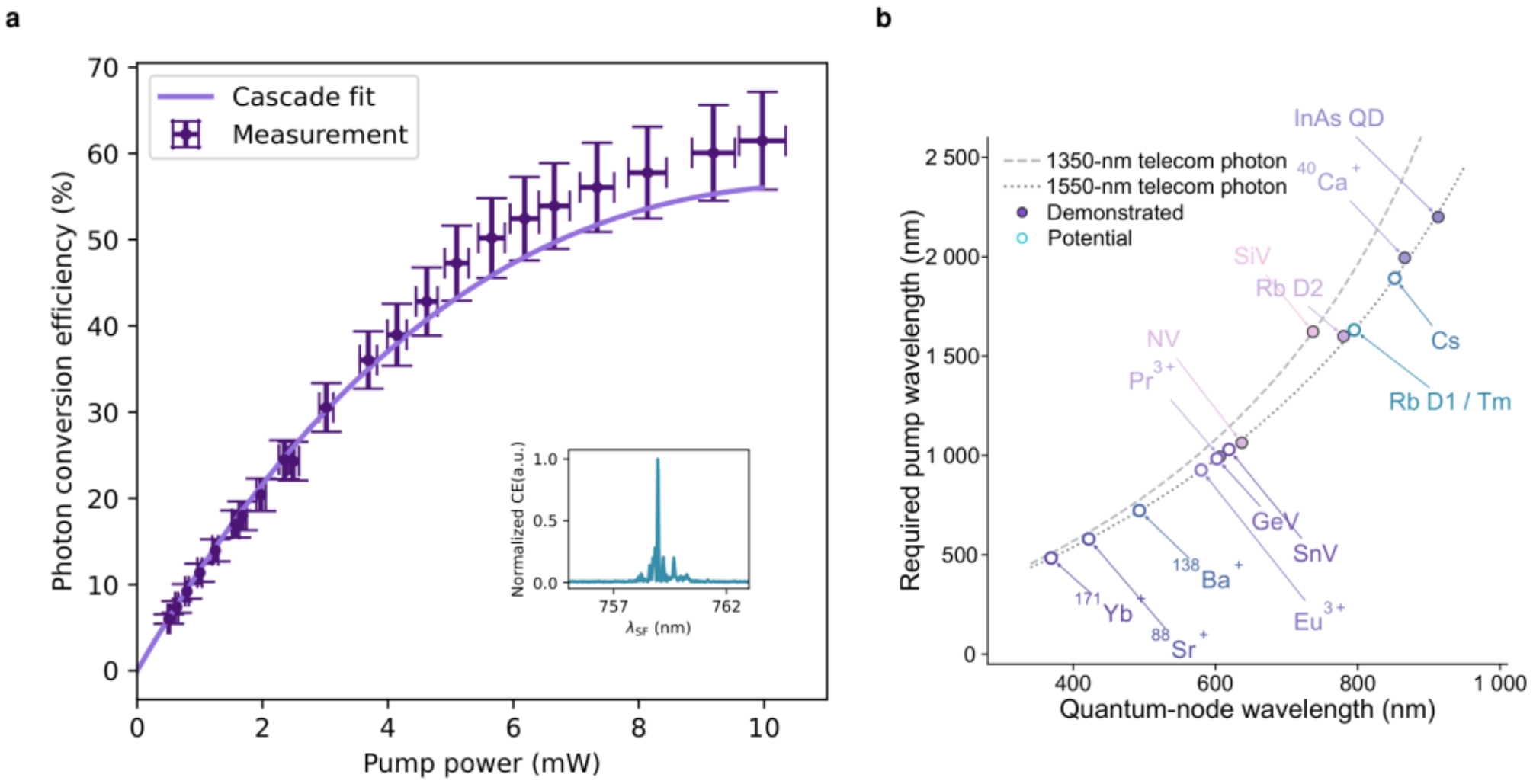

**a,** On-chip sum-frequency generation photon CE as a function of pump power. The pump and idler wavelengths are fixed at 1591.9 nm and 1450.5 nm, respectively, generating a signal near 758.9 nm. Purple markers show the measured on-chip CE, and the solid line shows a fit using an undepleted-pump with propagation and bend losses across the folded device (assuming identical efficiency for each section and ideal inter-section phase alignment). Error bars show the experimental uncertainty (see Supplementary Information). The inset shows the measured SFG spectrum for a fixed idler wavelength of 1450.5 nm. **b,** Pump wavelengths required for converting telecom photons at 1350 nm (dashed line) and 1550 nm (dotted line) to representative quantum-node transitions. Labeled markers identify selected solid-state emitters, rare-earth ions, trapped ions and atomic transitions. Filled markers denote wavelength combinations for which quantum frequency conversion has been experimentally demonstrated, whereas open markers indicate potential conversion interfaces.

## Methods

### Device fabrication

Both the folded TF-PPLN devices and the racetrack resonators were fabricated on commercially available 4-inch thin-film lithium niobate-on-insulator wafers (NanoLN). The nominal material stack consisted of a 300 nm x-cut, 5% MgO-doped lithium niobate device layer on a 7 μm buried oxide layer and a silicon handle. The devices were fabricated using an etch-then-pole with sidewall-poling process, with the film thickness characterized by ellipsometry both before and after etching. The waveguides were first patterned by electron-beam lithography using ma-N 2405 resist and transferred into the lithium niobate layer by dry etching to a depth of approximately 150 nm and waveguide width of 1.6 μm. This process defines the final waveguide geometry before poling, allowing the quasi-phase-matching design to account for the fabricated structure. After waveguide etching, poling electrodes were patterned adjacent to the waveguide for sidewall poling[46] In order to compensate for systematic deviations in the fabricated geometry and align the phase-matching wavelength of the PPLN sections to the target wavelength, the local poling period was selected using the adapted-poling with calibration procedure[2].

Periodic domain inversion was performed by applying high-voltage pulses to the poling electrodes. A pulse amplitude of approximately 225 V was used, with an effective pulse duration of approximately 1 ms and a total pulse sequence duration of approximately 1800 ms . After poling, the domain inversion quality was inspected using second-harmonic microscopy. The metal poling electrodes were then removed by wet etching.

Following waveguide fabrication and poling, the devices were cladded with 1.4 μm of $SiO_2$ deposited by plasma-enhanced chemical vapor deposition. Integrated thermo-optic heaters were subsequently fabricated on top of the oxide cladding. The heater stack consisted of 100 nm thick approximately 6.5 μm wide Pt for the resistive heater layer and 400 nm Au for electrical routing. Each 1-cm PPLN section was equipped with a peak-shifting heater to tune its local phase-matching wavelength. Additional heaters were placed between adjacent PPLN sections to tune the relative phase accumulated between the fundamental and second-harmonic fields in the connecting bends. The PPLN shifters have an average resistance of 342 Ω and the phase shifters have resistance of 188 Ω. Trenches were opened through the oxide cladding, lithium niobate layer, and buried oxide to improve thermal heating efficiency[50]. The chips were then diced at the facets for optical coupling. Finally, the exposed silicon handle was selectively undercut using $XeF_2$ etching to improve the thermal tuning efficiency and to remove the uneven Si handle from dicing. The cross section of the device and heater geometry are shown in Extended Fig. 6(i) and (ii). Since the available multichannel voltage source has a maximum output voltage of 18 V, the integrated heaters were designed with a parallel geometry, as shown in Extended Fig. 6(iii), to reduce their electrical resistance and allow sufficient thermal tuning power within this voltage range.

### Bend Loss and Propagation Loss Characterization

Bending loss was characterized using a series of racetrack resonators with identical pulley couplers and Euler bends but different straight-section lengths. The racetrack waveguide width was 1.6 μm, matching that of the PPLN waveguides, while the bus waveguide in the pulley coupler was 0.5 μm wide with a center-to-center separation of 1.05 μm. The pulley coupler was kept identical for all devices and designed to operate near critical coupling at both 1600 nm and 780 nm. The bend sections were implemented using 1.6 μm-wide Euler bends. Integrated heaters with the same metal stack as in the folded PPLN devices were included for thermal tuning. The straight-section lengths were varied as 3400, 2600, 1800, 1000, and 400 μm. For each resonator, the measured transmission resonances were fitted to extract the intrinsic and external quality factors, $Q_i$ and $Q_e$, as well as the free spectral range. The intrinsic round-trip loss was calculated from $\alpha_{roundtrip} = \frac{10}{\ln(10)} \cdot \frac{2\pi \cdot c}{Q_i \cdot FSR \cdot \lambda_0}$, where $\lambda_0$ is the optical wavelength. As shown in Extended Figure 3, the extracted round-trip loss was then plotted against the total straight-section length and fitted with a linear model. One device showing clear mode-crossing signatures, including side peaks or non-Lorentzian line shapes near the main resonance, was excluded from the propagation-loss fit at visible wavelength because the extracted intrinsic linewidths are not reliable under a single-resonance Lorentzian model. One additional device was excluded from the fit at telecom wavelength because its extracted round-trip loss was inconsistent with the linear length trend and strongly affected the regression. Excluded devices are shown in gray but were not included in the linear fit. The slope gives the straight-waveguide propagation loss, while the intercept at zero straight-section length gives the fixed round-trip excess loss from the two Euler bends and the coupler region. Assuming negligible or length-independent coupler excess loss, the bend and coupling losses were extracted from the zero-length intercept and the bend loss was estimated to be one half of this value. This value represents an upper bound on the per-bend loss unless the pulley-coupler excess loss is independently calibrated.

**SHG section characterization and estimation of optimal efficiency**

To estimate the effective normalized CE of the folded device from isolated-section measurements, we first calibrated the measured powers to the local reference plane of each PPLN section. For the j-th section, the locally measured normalized efficiency is defined as

$$\eta_{j,local}^{lossy} = 100\% \cdot \, P_{2\omega,j}^{exit} / (P_{\omega,j}^{in})^2,$$

where $P_{\omega,j}^{in}$ is the fundamental power at the input of the selected PPLN section and $P_{2\omega,j}^{exit}$ is the second-harmonic power immediately after that section. The measured powers were corrected for the independently measured propagation and bend losses between each section and the input/output value. For PPLN1, PPLN2, and PPLN3, the second-harmonic output corrections correspond to propagation through 2 cm + 2 bends, 1 cm + 1 bend, and 0 cm + 0 bends, respectively. The corresponding fundamental attenuation corrections from the chip input to each section input are 0 cm + 0 bends, 1 cm + 1 bend, and 2 cm + 2 bends, respectively. This procedure gives the local section efficiency referenced to the physical input and output of each active PPLN section, as specified in Supplementary Information.

Because these locally extracted efficiencies still include propagation loss within the active 1-cm PPLN section, we convert them to equivalent lossless section efficiencies before using them in the lossy folded coupled-mode model. In the undepleted and phase-matched limit, the fundamental field inside one section evolves as

$$a_\omega(z) = a_\omega(0)e^{-\alpha_\omega z/2},$$

and the second-harmonic field generated at the section output is proportional to

$$a_{2\omega}(L) \propto \int_0^L \kappa\, a_\omega^2(0)\, e^{-\alpha_\omega z-\alpha_{2\omega}(L-z)/2}dz = \kappa\, a_\omega^2(0)\, e^{-\alpha_{2\omega}L/2}\, \frac{1- e^{-(\alpha_\omega-\alpha_{2\omega}/2)L}}{\alpha_\omega-\alpha_{2\omega}/2}.$$

Here, $|a_\omega(z)|^2 = P_\omega(z)$ and $|a_{2\omega}(z)|^2 = P_{2\omega}(z)$.

Since for the lossless case, $a_{2\omega,\mathrm{lossless}}(L) = \kappa\, a_\omega^2(0)\, L$, we define an intra-section loss reduction factor

$$F_{\mathrm{loss}} = \left( e^{-\alpha_{2\omega}L/2}\, \frac{1- e^{-(\alpha_\omega-\alpha_{2\omega}/2)L}}{\alpha_\omega-\alpha_{2\omega}/2}\, \frac{1}{L}\right)^2.$$

The attenuation coefficients are converted from the measured propagation losses by $\alpha_j = \frac{\ln 10}{10} 100 \cdot \alpha_{dB/cm,j}$, where $\alpha_j$ is in $m^{-1}$. Using $L = 1cm, \alpha_{dB/cm,\omega} = 0.18\ dB/cm, \alpha_{dB/cm,2\omega} = 1.30\ dB/cm$, we obtain $F_{\mathrm{loss}} \approx 0.827$. The equivalent lossless nonlinear efficiency used in the coupled-mode simulation is therefore $\eta_j^{\mathrm{lossless}} = \eta_j^{\mathrm{lossy}}/F_{\mathrm{loss}}$ and is then used as the nonlinear coupling strengths in the full folded three-PPLN model. For the lossless case, $F_{\mathrm{loss}} = 1$.

**Simulation of the folded 3-section PPLN**

In this model, each 1-cm PPLN section is treated as a lossy nonlinear waveguide with its independently extracted intrinsic efficiency $\eta_j^{\mathrm{lossless}}$, while propagation losses at the fundamental and second-harmonic wavelengths are applied continuously along each section. The losses introduced by the connection bends for the fundamental and second-harmonic fields have been accounted for. The field evolution is therefore modeled as

$$\frac{da_\omega}{dz} = -\frac{\alpha_\omega}{2} a_\omega - i\kappa_j^* \cdot a_\omega^* \cdot a_{2\omega},$$

$$\frac{da_{2\omega}}{dz} = -\frac{\alpha_{2\omega}}{2} a_{2\omega} - i\kappa_j \cdot a_\omega^2,$$

where $\kappa_j = \frac{\sqrt{\eta_j^{\mathrm{lossless}}}}{L_j}$ uses the numerical section efficiency in $W^{-1}$ inside the square root (divide a value quoted in %/W by 100) and is chosen such that, in the lossless and undepleted limit, the j-th section gives the normalized CE $\eta_j^{\mathrm{lossless}}$. Between adjacent PPLN sections, the fundamental and second-harmonic fields are multiplied by the corresponding bend-loss factors and accumulate a relative phase controlled by the inter-PPLN heaters. The heaters are tuned so that the second-harmonic fields generated in all three sections add constructively at the output. The effective folded normalized CE is then extracted from the simulated input–output response as

$$\eta_{cas} = 100\% \cdot P_{2\omega,\mathrm{out}}/P_{\omega,\mathrm{in}}^2$$

in the low-conversion, undepleted-pump regime. Here $P_{\omega,\mathrm{in}}$ is the fundamental power at the input of the folded device and $P_{2\omega,\mathrm{out}}$ is the second-harmonic power at the chosen output reference plane. Because the propagation and bend losses are already included explicitly in the folded model, $\eta_{\mathrm{cas}}$ represents the effective device-level overall CE, while the fitted $\eta_j^{\mathrm{lossless}}$ values represent the intrinsic nonlinear efficiencies of the individual PPLN sections.

**Tuning optimization**

The folded device was optimized through an iterative heater-tuning procedure, as summarized in Extended Fig. 4. The optimization alternated between tuning the peak-shifting heaters on the PPLN sections and the inter-PPLN phase shifters. For a given PPLN section, the corresponding PPLN heater was first scanned to align its phase-matching response to the target wavelength. The two inter-PPLN phase shifters were then scanned sequentially to maximize the coherently combined SH output. Each heater scan was performed by recording the SH power at the target wavelength as a function of the applied heater power, and the heater power giving the largest SH signal was selected for the next step.

The optimization was carried out in coarse and fine stages. In the coarse stage, larger heater-power steps were used to bring the three PPLN phase-matching peaks close to the target wavelength and to approximately set the inter-PPLN phases. The same sequence was then repeated with smaller heater-power steps for fine optimization. The tuning sequence consisted of coarse tuning of PPLN1 followed by phase shifters 1 and 2, fine tuning of PPLN1 followed by phase shifters 1 and 2, and the same coarse/fine procedure for PPLN2 and PPLN3. Selected tuning steps from this procedure are shown in Extended Fig. 4, where the orange markers indicate the heater powers chosen after each scan.

After each tuning step, the SHG spectrum was measured and the peak overall CE was used to monitor the optimization progress. The procedure was stopped after 22 tuning steps, when additional scans of the PPLN and phase-shifter heaters did not produce a further increase in the peak CE.

**SFG normalization and fitting**

The photon conversion is defined as the ratio of the output visible photon flux to the input idler photon flux, and power conversion is defined as the corresponding ratio of optical powers. The photon-conversion fraction is $\Gamma = \frac{P_{\mathrm{vis,out}}}{P_{\mathrm{idler,in}}}\frac{\lambda_{\mathrm{vis}}}{\lambda_{\mathrm{idler}}}$, so the power-conversion fraction is $\frac{P_{\mathrm{vis,out}}}{P_{\mathrm{idler,in}}} = \Gamma\frac{\lambda_{\mathrm{idler}}}{\lambda_{\mathrm{vis}}}$. At $\lambda_{\mathrm{idler}} = 1450.5$ nm and $\lambda_{\mathrm{vis}} = 758.9$ nm, power conversion is approximately 1.911 times photon conversion. Each facet transmission in decibels is expressed as $L_{\mathrm{facet,x}} = 10\log_{10}(T_{\mathrm{facet,x}})$, using a separate calibration for each wavelength and measurement session. The on-chip powers are obtained by dividing by the corresponding facet transmissions, as described in the Supplementary Information. The resulting initial photon-conversion slope is approximately 12,599 %/W. Multiplying by $\frac{\lambda_{\mathrm{idler}}}{\lambda_{\mathrm{vis}}} \approx 1.911$ gives a normalized power-conversion efficiency of 24,078 ± 656 %/W, accounting for propagation and bend losses in the device. The full derivation of the SFG conversion efficiency and its uncertainty is provided in the Supplementary Information.

## Extended Data

**Extended Data Figure 1: Characterization of crosstalk between adjacent thermally-tuned PPLN waveguides.**

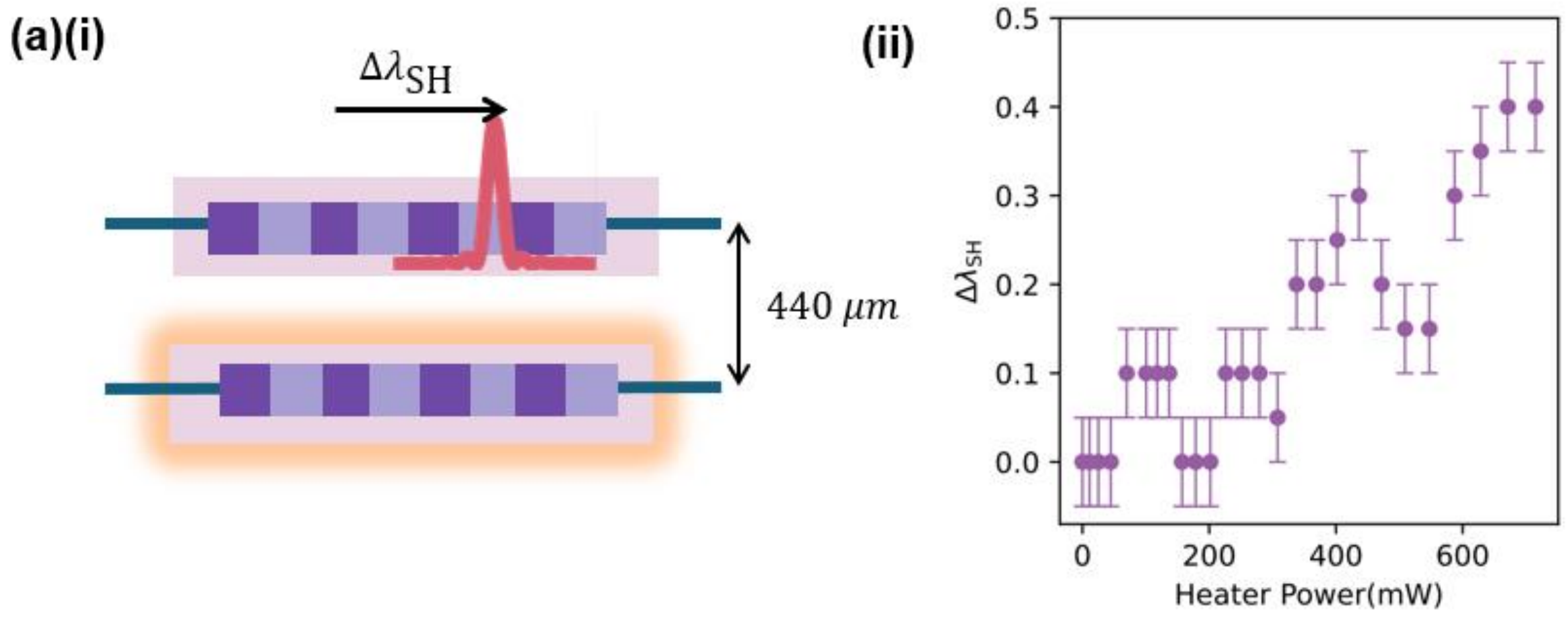

**a.(i)** Two adjacent PPLN sections, separated by approximately 440 µm, were used to quantify thermal crosstalk. Electrical power was applied to the heater of one PPLN section, while the phase-matching peak wavelength of the neighboring PPLN was monitored. **(ii)**, The measured peak shift $\Delta\lambda_{SH}$ remains negligible for heater powers below ~200 mW, whereas a finite wavelength shift is observed at higher powers, indicating measurable thermal crosstalk between the two sections.

**Extended Data Figure 2: Single PPLN Heating Efficiency**

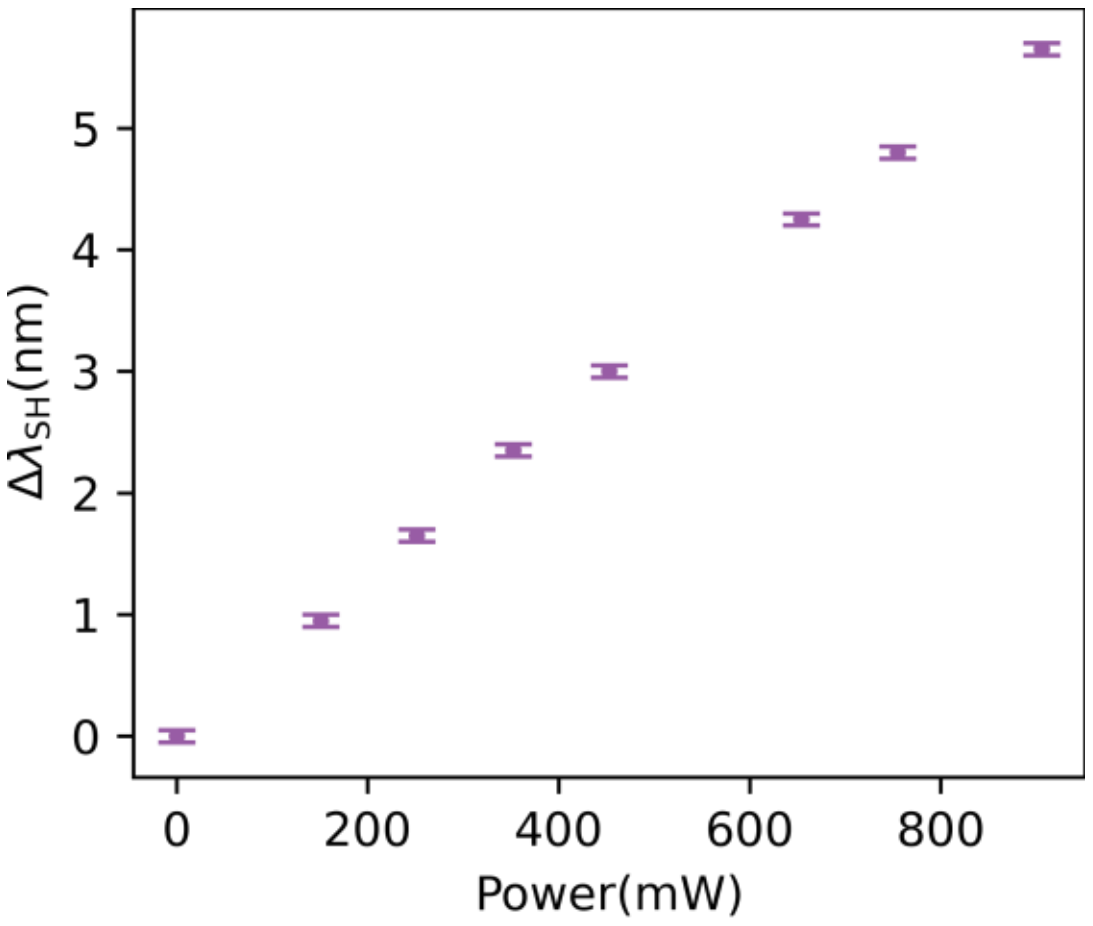

Measured thermo-optic tuning of the quasi-phase-matching peak wavelength in a single 1.0-cm-long PPLN section. The graph illustrates the second-harmonic wavelength shift, $\Delta\lambda_{SH}$, as a function of the electrical power applied to the integrated peak-shifting heater. This measurement is used to determine the heating efficiency of the individual PPLN sections.

**Extended Data Figure 3: Bend-loss and Propagation loss**

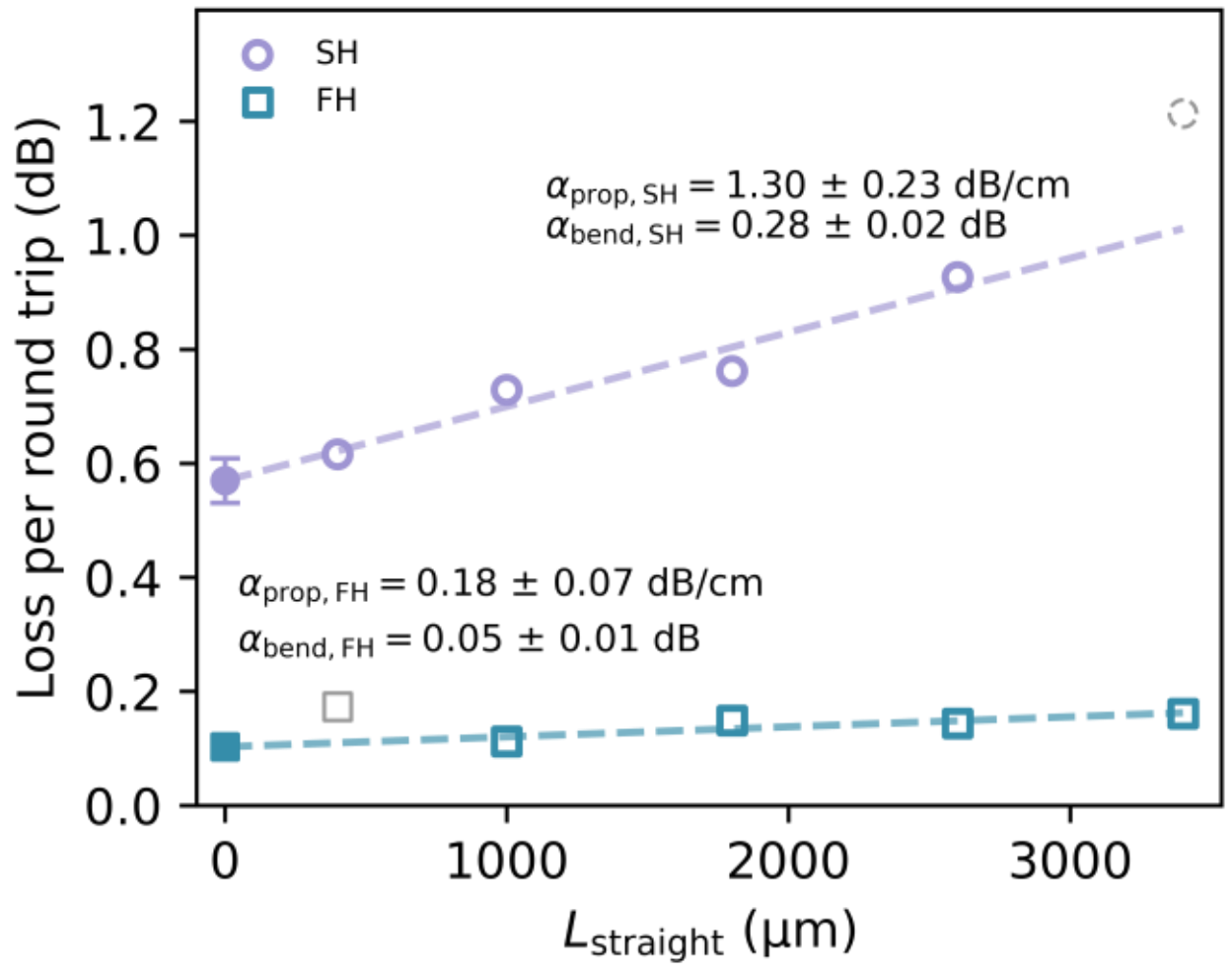

Round-trip loss as a function of the straight connecting length $L_{straight}$, including both bend loss, $\alpha_{bend}$, and the straight-waveguide propagation loss $\alpha_{prop}$. Purple circles (SH, near 780 nm) and cyan squares (FH, near 1600 nm) show measured round-trip loss. The fitted per-bend losses are 0.28 ± 0.02 and 0.05 ± 0.01 dB/bend, respectively. The extracted propagation loss is $\alpha_{prop,FH} = 0.18 \pm 0.07$ dB/cm at 1600 nm and $\alpha_{prop,SH} = 1.30 \pm 0.23$ dB/cm at 780 nm. Two devices are excluded due to mode-crossing signatures and inconsistency with the linear length; these data points are shown in gray but are not included in the linear fit.

## Extended Data Figure 4: Heater-power scans during iterative optimization

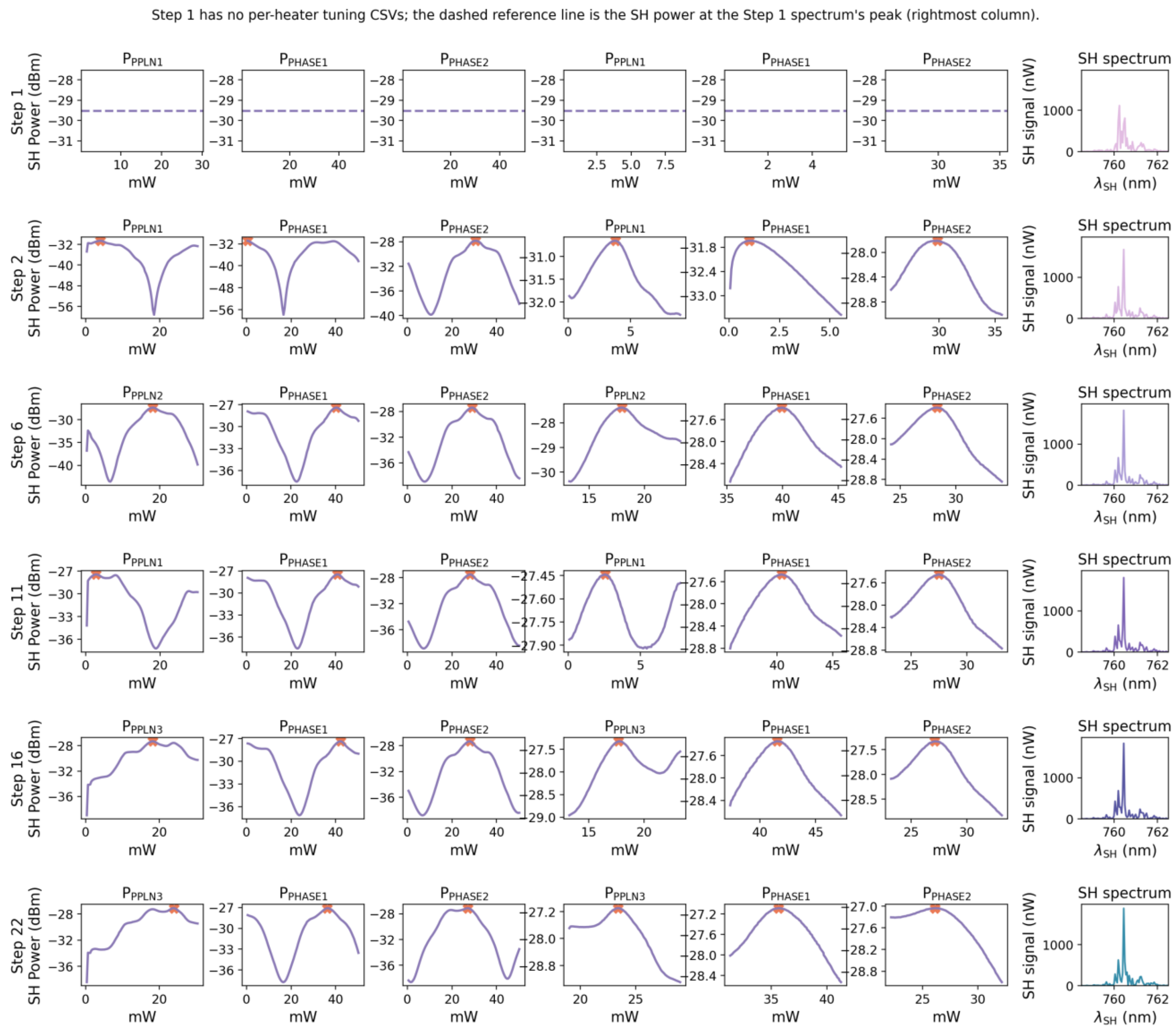

Measured second-harmonic (SH) power as a function of the electrical power applied to each integrated heater for selected optimization steps corresponding to Fig. 3b(i). Each row shows one optimization step, and the six panels in each row show the coarse and refined heater scans used to optimize the selected PPLN peak-shifting heater and the two inter-PPLN phase shifters. The horizontal axis is the applied heater power, and the vertical axis is the measured SH power in dBm. Orange crosses mark the heater powers selected at each scan step. The seventh panel in each row shows the measured SH spectrum in nW. No per-heater scan data are available for Step 1; the dashed lines in its first six panels are reference levels derived from the measured Step-1 spectrum and are not heater scans.

**Extended Data Figure 5: Measurement Setup**

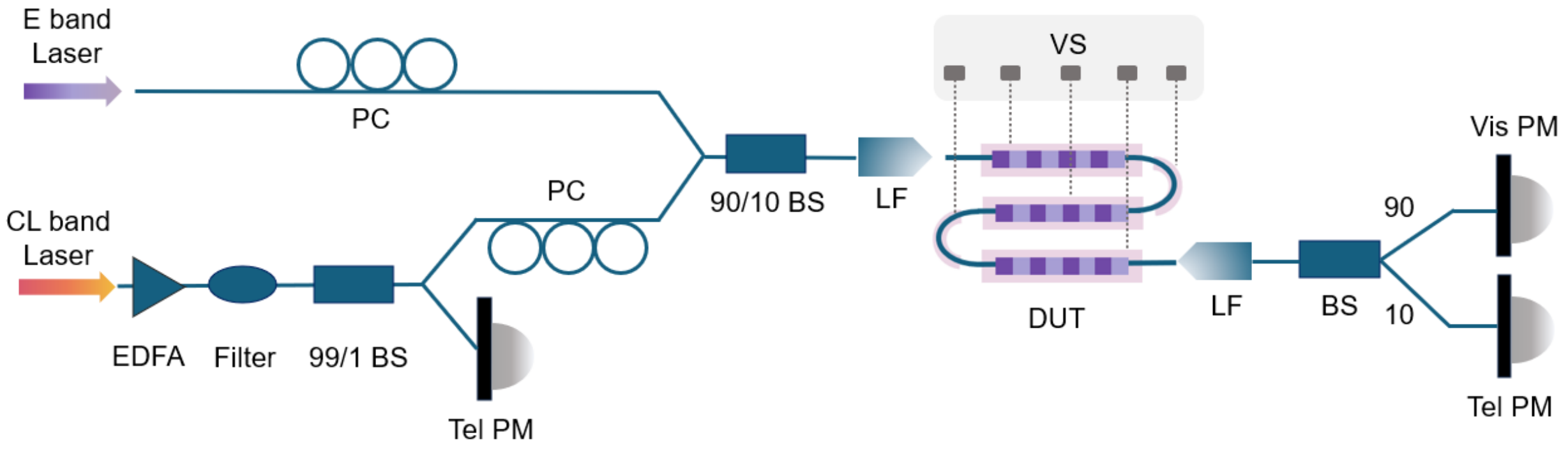

E-band: Santec TSL-550; 1355 nm to 1485 nm; CL-band: Santec TSL-710; 1480 nm to 1640 nm; PC, polarization controller; EDFA: amplifier (AEDFA-L-33); Filter: tunable filter (XTM-50); BS: beam splitter; LF: lensed fiber; DUT: device under test; Tel/Vis PM: telecom/visible powermeter; VS: multichannel voltage source (Nicslab, XDAC-8MUB-R4G8-SMA)

**Extended Data Figure 6: Cross-sectional schematic of the thermally isolated folded TF-PPLN waveguide.**

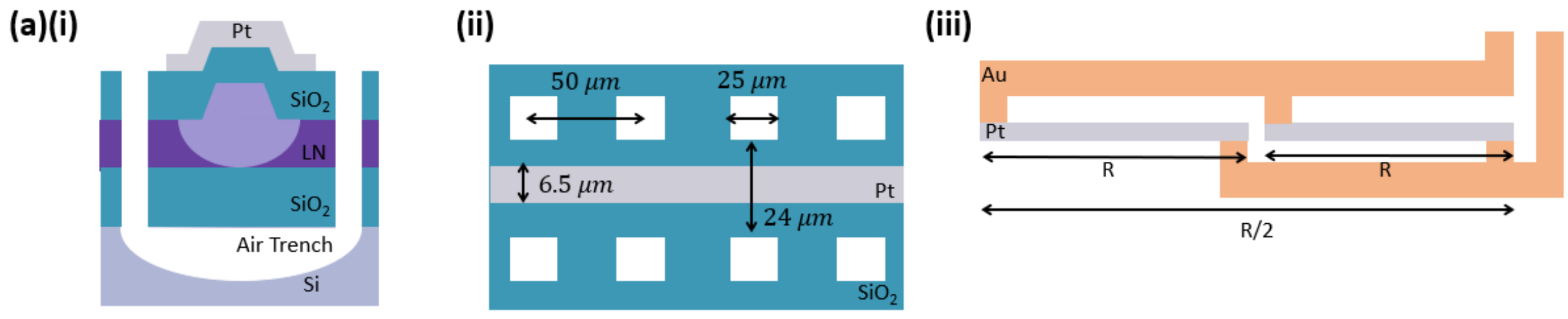

**(i)** Cross-sectional schematic of the waveguide and heater stack. The device is fabricated from a 300-nm-thick x-cut, 5% MgO-doped lithium niobate layer on a 7-µm-thick buried $SiO_2$ layer and silicon handle. The lithium niobate waveguide is etched by approximately 150 nm and clad with 1.4 µm of PECVD $SiO_2$, which separates the optical mode from the integrated Pt heater. Trenches are opened through the top oxide, lithium niobate layer, and buried oxide, followed by $XeF_2$ undercut of the exposed silicon to form an air trench beneath the waveguide for improved thermal isolation. **(ii)** Top-view schematic of the parallel heater layout used to reduce electrical resistance and provide sufficient tuning power within the 18-V output limit of the multichannel voltage source. The Pt heater consists of 6.5-µm-wide resistive sections connected in parallel, with 50-µm-long heater segments, 25-µm-wide openings, and a 24-µm spacing between the parallel arms. **(iii)** Parallel heater geometry used to reduce the electrical resistance of a long thermo-optic heater. The Au routing layer connects two Pt heater segments in parallel, each with resistance (R), giving an effective resistance of (R/2) while maintaining the full heater length.

**Extended Data Table 1: Optimized electrical powers applied to the three PPLN peak-shifting heaters and two inter-PPLN phase-shifting heaters.**

| | $P_{Phase1}$ | $P_{Phase2}$ | $P_{PPLN1}$ | $P_{PPLN2}$ | $P_{PPLN3}$ |
|---|---|---|---|---|---|
| Power (mW) | 35.6 | 26.1 | 5.1 | 17.8 | 23.3 |

**Data availability** The data that support the plots within this paper and other findings of this study are available from the corresponding author upon reasonable request.

**Code availability** The code used to produce the plots within this paper is available from the corresponding author upon reasonable request.

**Acknowledgements**

We thank Rebecca Cheng and Hana Warner for discussions. This work was supported by the Defense Advanced Research Projects Agency (HR0011-26-9-E125, HR0011-26-9-E119, HR0011-24-2-0360); A*STAR National Science Scholarship; National Research Foundation of Korea (NRF-2022M3K4A1094782); AWS Center for Quantum Networking's research alliance with the Harvard Quantum Initiative; Harvard University Dean's Competitive Fund for Promising Scholarship; Harvard Quantum Initiative; NASA (80NSSC22K0262, 80NSSC23PB442); NSF Engineering Research Center for Quantum Networks (EEC-1941583); NSF QuiC-TAQS (2137723); IonQ Inc.; MagiQ Technologies: Naval Air Warfare Center Aircraft Division(N6833522C0413); the Dutch Research Council Rubicon postdoctoral fellowship (019.231EN.011); and the Gordon and Betty Moore Foundation. Device fabrication was performed at the Center for Nanoscale Systems (CNS), a member of the National Nanotechnology Coordinated Infrastructure Network (NNCI), which is supported by the National Science Foundation under NSF Grant No. 1541959.

**Author Contributions**

S.L. designed and fabricated the devices, and completed experiments with assistance and advice from C.X. and J.Y. C.X. and A.S. proposed the idea. A.C. and Y.S. assisted with resonator design and loss measurement. C.A.A.F. and Y.H. assisted with heater design. X.Z. and L.M. helped develop the fabrication process. M.L. supervised the project. S.L. and M.L. wrote the manuscript in collaboration with the remaining co-authors.

**Competing Interests**

M.L. is involved in developing lithium niobate technologies at HyperLight Corporation. C.A.A.F. is involved in developing lithium niobate technologies at Sabratha Photonics B.V. The remaining authors declare no competing interests.

# Supplementary

**Uncertainty estimation for SHG and SFG efficiencies**

The uncertainties of the SHG and SFG efficiencies were evaluated by propagating the calibration uncertainties in logarithmic units. All optical powers were first corrected to the corresponding on-chip values using the measured facet, propagation, and bend losses. P denotes linear optical power in watts. Correction terms are included explicitly as $B_{\omega}$ or $B_{2\omega}$ (in dB) and account for the beam-splitter ratio when converting the measured detector power to the power in the optical path. A small uncertainty δL in a loss correction expressed in dB gives a relative power uncertainty of

$$\delta P / P = (\ln 10 / 10)\, \delta L.$$

Therefore, all uncertainties were first propagated in dB units and then converted to fractional uncertainties by multiplying by ln 10 / 10.

The telecom and visible propagation losses are denoted as $\alpha_{\mathrm{prop,tel}}$ and $\alpha_{\mathrm{prop,vis}}$, respectively, in units of dB/cm. The telecom and visible bend losses are denoted as $\alpha_{\mathrm{bend,tel}}$ and $\alpha_{\mathrm{bend,vis}}$, respectively, in units of dB/bend. Their one-standard-deviation uncertainties are denoted as $\sigma_{\alpha_{\mathrm{prop,x}}}$ and $\sigma_{\alpha_{\mathrm{bend,x}}}$, where $\mathrm{x} = \mathrm{tel\ or\ vis}$.

Because the facet losses were extracted from reference transmission measurements after subtracting propagation and bend losses, the propagation- and bend-loss uncertainties contribute to the facet-loss uncertainty. For a reference structure with measured total loss $\mathrm{L_{tot,ref}}$, propagation length $\mathrm{l_{ref}}$, and number of bends $\mathrm{N_{bend}}$, the per-facet loss at wavelength band x is

$$\mathrm{L_{facet,x}} = -(\mathrm{L_{tot,ref,x}} - \alpha_{\mathrm{prop,x}}\mathrm{l_{ref}} - \alpha_{\mathrm{bend,x}}\mathrm{N_{bend,x}})/2\,,$$

Here $\mathrm{L_{facet,x}} = 10\, log_{10}(\mathrm{T_{facet,x}}) \leq 0$ is the signed facet transmission; propagation and bend losses remain positive attenuation magnitudes. Each wavelength and measurement session uses its own calibration.

The corresponding facet-calibration uncertainty is

$$\sigma^2_{\mathrm{L_{facet,x}}} = \frac{1}{4}\left( \sigma^2_{\mathrm{L_{tot,ref,x}}} + \mathrm{l^2_{ref,x}}\, \sigma^2_{\alpha_{\mathrm{prop,x}}} + \mathrm{N^2_{bend,x}}\, \sigma^2_{\alpha_{\mathrm{bend,x}}}\right).$$

Here, $\sigma^2_{\mathrm{L_{tot,ref,x}}}$ is the uncertainty of the measured reference transmission. Assuming 0.1 dB uncertainty for each independent optical power measurement, the uncertainty of a transmission measurement obtained from the difference between input and output powers is $\sigma_{\mathrm{L_{tot,ref,x}}} = \sqrt{2} \cdot 0.1$ dB. Both the telecom and visible reference structures are modeled as 3 cm long with two bends, matching the central facet-loss calibration. Including these two bends in the uncertainty calculation gives a full-device SHG result of 15777 ± 1776 %/W under the stated calibration assumptions.

**SHG uncertainty**

The overall SHG CE is defined using the on-chip fundamental input power and the on-chip second-harmonic output power:

$$\eta_{SHG} = 100\ P_{2\omega,on-chip}/P^2_{\omega,on-chip}.$$

For the full device, $I_{tel} = \alpha_{prop,tel} l_{dev} + \alpha_{bend,tel} N_{dev}$ with $l_{dev} = 3\ cm$ and $N_{dev} = 2$. The SHG correction factors are $B_\omega$ and $B_{2\omega}$. Thus

$$P_{\omega,on-chip} = P_{\omega,meas} \times 10^{\frac{B_\omega - L_{facet,tel} + I_{tel}}{10}}.$$

The on-chip second-harmonic output power is reconstructed as:

$$P_{2\omega,on-chip} = P_{2\omega,meas} \times 10^{\frac{B_{2\omega} - L_{facet,vis}}{10}}.$$

Therefore,

$$\eta_{SHG} = 100\ \frac{P_{2\omega,meas}}{P_{\omega,meas}{}^2} \times 10^{\frac{B_{2\omega} - 2B_\omega - L_{facet,vis} + 2L_{facet,tel} - 2I_{tel}}{10}}.$$

The uncertainty of the full-device SHG efficiency is

$$\sigma_{\eta_{SHG}} = \eta_{SHG} \cdot \frac{\ln 10}{10}\left(\sigma^2_{P_{2\omega}} + 4\sigma^2_{P_\omega} + \sigma^2_{L_{facet,vis}} + 4\sigma^2_{L_{facet,tel}}\right)^{1/2}.$$

The facet and internal-path corrections are correlated because the same internal losses enter the facet calibration. For, $Cov(L_{facet,tel}, I_{tel}) = \frac{\sigma_{I,tel}{}^2}{2}$, so $Var(2L_{facet,tel} - 2I_{tel}) = 4\,Var(L_{facet,tel})$.

**Isolated PPLN-section SHG efficiency**

For the isolated PPLN-section measurements, the measured powers were additionally corrected to the local input and output planes of each selected PPLN section. For section j, the local SHG efficiency is

$$\eta_j = 100\ P_{2\omega,j,local,out}/P^2_{\omega,j,local,in}.$$

Using the measured detector powers and the same correction factors, the local efficiency is

$$\eta_j = 100\ \frac{P_{2\omega,meas}}{P_{\omega,meas}{}^2} \times 10^{\frac{B_{2\omega} - 2B_\omega + C_{vis,j} - 2C_{tel,j}}{10}}.$$

The reconstruction corrections $C_{tel,j}$ and $C_{vis,j}$ refer to the path from section j's fundamental input to the telecom detector and from its second-harmonic output to the visible detector. For sections 1, 2 and 3, the telecom paths are 3, 2 and 1 cm with 2, 1 and 0 bends; the visible paths are 2, 1 and 0 cm with 2, 1 and 0 bends.

For wavelength channel x, the reconstruction correction is

$$C_{\mathrm{x,j}} = -L_{\mathrm{facet,x}} + \alpha_{\mathrm{prop,x}} l_{\mathrm{x,j}} + \alpha_{\mathrm{bend,x}} N_{\mathrm{x,j}}.$$

Substituting the signed facet-calibration expression before propagating uncertainties gives

$$C_{\mathrm{x,j}} = \frac{L_{\mathrm{tot,ref,x}}}{2} + \alpha_{\mathrm{prop,x}}(l_{\mathrm{x,j}} - \frac{l_{\mathrm{ref,x}}}{2}) + \alpha_{\mathrm{bend,x}}(N_{\mathrm{x,j}} - \frac{N_{\mathrm{ref,x}}}{2}).$$

Therefore, the uncertainty of the local correction is

$$\sigma_{\mathrm{C,x,j}}{}^2 = \frac{\sigma_{\mathrm{L,tot,ref,x}}{}^2}{4} + (l_{\mathrm{x,j}} - \frac{l_{\mathrm{ref,x}}}{2})^2 \sigma_{\alpha,\mathrm{prop,x}}{}^2 + (N_{\mathrm{x,j}} - \frac{N_{\mathrm{ref,x}}}{2})^2 \sigma_{\alpha,\mathrm{bend,x}}{}^2.$$

The uncertainty of the isolated-section SHG efficiency is

$$\sigma_{\eta,\mathrm{j}} = \eta_{\mathrm{j}} \frac{ln\,10}{10} (\sigma_{\mathrm{P,2\omega}}{}^2 + 4\sigma_{\mathrm{P,\omega}}{}^2 + \sigma_{\mathrm{C,vis,j}}{}^2 + 4\sigma_{\mathrm{C,tel,j}}{}^2)^{\frac{1}{2}}.$$

**Zero-loss limit of the three-section cascade**

The ideal zero-loss cascade efficiency in the main text, $\eta_{\mathrm{lossless}} \approx 34392\ \%/\mathrm{W}$, is the exact zero-loss limit of the same validated three-section coupled-mode model used elsewhere in this work. Each section's own independently measured local (lossy) normalized efficiency, $\eta_{\mathrm{j,local}}$ (sections 1, 2 and 3 are not identical, owing to fabrication and calibration variation), is corrected for intra-section propagation loss using the factor $F_{\mathrm{loss}}$ (1 cm) defined in the main-text Methods, giving a lossless-equivalent coefficient $\eta_{\mathrm{j,lossless}} = \eta_{\mathrm{j,local}} / F_{\mathrm{loss}}(1\ \mathrm{cm}) \approx 5004, 3126$ and $3458\ \%/\mathrm{W}$ for sections 1, 2 and 3. In the lossless, undepleted-pump limit, the coupled-mode equations for the folded cascade integrate exactly to coherent field-amplitude addition,

$$\eta_{\mathrm{lossless}} = (\sqrt{\eta_{1,\mathrm{lossless}}} + \sqrt{\eta_{2,\mathrm{lossless}}} + \sqrt{\eta_{3,\mathrm{lossless}}})^2,$$

so the reported value coherently sums the three unequal per-section amplitudes rather than assuming identical sections. This is a model extrapolation using the calibrated section coefficients, not a measured zero-loss efficiency.

**Pump-power uncertainty for SFG**

For the SFG measurement, the CE is plotted as a function of the on-chip pump power. Since the plotted vertical axis is photon CE from idler to visible, the pump-power uncertainty is propagated separately as a horizontal error bar.

The on-chip pump is

$$P_{\mathrm{pump,on}} = P_{\mathrm{pump,meas}} \times 10^{\frac{B_{\mathrm{pump}} - L_{\mathrm{facet,pump}}}{10}}.$$

The absolute pump-power uncertainty is

$$\sigma_{\mathrm{P,pump,on}} = \frac{ln\,10}{10} P_{\mathrm{pump,on}} (\sigma_{\mathrm{P,pump}}{}^2 + \sigma_{\mathrm{L,facet,pump}}{}^2)^{\frac{1}{2}}.$$

**SFG photon CE**

The SFG photon CE is defined as the ratio between the generated visible photon flux and the input idler photon flux. Since photon flux is power divided by photon energy, this can be written as

$$\Gamma_{\mathrm{SFG}} = 100\,\frac{\mathrm{P_{vis,on}}}{\mathrm{P_{idler,on}}}\frac{\lambda_{\mathrm{vis}}}{\lambda_{\mathrm{idler}}}.$$

For SFG, the visible detector reading is corrected using the signed visible facet transmission and correction factor $B_{\mathrm{vis}} = 1.50\ dB$:

$$P_{\mathrm{vis,on-chip}} = P_{\mathrm{vis,meas}} \times 10^{\frac{B_{\mathrm{vis}} - L_{\mathrm{facet,vis}}}{10}}.$$

And the on-chip idler power is

$$P_{\mathrm{idler,on-chip}} = P_{\mathrm{idler,meas}} \times 10^{\frac{B_{\mathrm{idler}} + L_{\mathrm{facet,idler}}}{10}}.$$

Therefore,

$$\Gamma_{\mathrm{SFG}} = 100\,\frac{P_{\mathrm{vis,meas}}}{P_{\mathrm{idler,meas}}}\frac{\lambda_{\mathrm{vis}}}{\lambda_{\mathrm{idler}}} \times 10^{\frac{B_{\mathrm{vis}} - B_{\mathrm{idler}} - L_{\mathrm{facet,vis}} - L_{\mathrm{facet,idler}}}{10}}.$$

Under the assumption of independent input quantities, the uncertainty of the SFG photon CE is

$$\sigma_{\Gamma,\mathrm{SFG}} = \Gamma_{\mathrm{SFG}}\frac{ln\,10}{10}(\sigma_{\mathrm{P,vis}}{}^2 + \sigma_{\mathrm{P,idler}}{}^2 + \sigma_{\mathrm{L,facet,vis}}{}^2 + \sigma_{\mathrm{L,facet,idler}}{}^2)^{\frac{1}{2}}.$$

The power-reading uncertainties are in dB.